%% file: main.tex
\documentclass[preprint,12pt]{elsarticle}

\usepackage{amssymb}
\usepackage{amsmath}

\usepackage{subcaption}
\usepackage{caption}

\usepackage{multirow}
\usepackage{soul} 
\usepackage{wrapfig}
\usepackage{float}
\usepackage{physics}
\usepackage{hyperref}

 \usepackage{lineno}
\date{}
\usepackage{geometry}

\newgeometry{top=1in,bottom=0.8in,right=1.0in,left=1.0in}

\journal{Nuclear Instruments and Methods A}

\begin{document}

\begin{frontmatter}



\title{Measurement of the LCLS-II dark current using the LDMX Trigger Scintillator Prototype}

\author[label4]{Elizabeth Berzin}
\author[label2]{Lene Kristian Bryngemark}
\author[label5]{Robert Craig Group}
\author[label1]{Joesph Kaminski}
\author[label3]{Timothy Nelson}
\author[label4]{Rory O'Dwyer}
\author[label5]{Jessica Pascadlo}
\author[label3]{Emrys Peets}
\author[label3]{Benjamin Reese}
\author[label4]{Lauren Tompkins}
\author[label5]{Kieran Wall}
\author[label1]{Andrew Whitbeck}

\affiliation[label1]{organization={Fermi National Accelerator Laboratory},
            city={Batavia},
            postcode={60510},
            state={IL},
            country={USA}}

\affiliation[label2]{organization={Lund University, Department of Physics},
            addressline={Box 118},
            city={Lund},
            postcode={221 00},
            country={Sweden}}

\affiliation[label3]{organization={SLAC National Accelerator Laboratory},
            city={Menlo Park},
            postcode={94025},
            state={CA},
            country={USA}}

\affiliation[label4]{organization={Stanford University},
            city={Stanford},
            postcode={94305},
            state={CA},
            country={USA}}

\affiliation[label5]{organization={University of Virginia},
            city={Charlottesville},
            postcode={22904},
            state={VA},
            country={USA}}

\begin{abstract}

\input{abstract.tex}
\end{abstract}







\end{frontmatter}

\input{intro.tex}

\input{detector.tex}

\input{results.tex}

\input{conclusion.tex}
\section*{Acknowledgements}
We would like to thank the SLAC engineers, technicians and  physicists who made this work possible: Mei Bai, Alev Ibrahimov, Thomas Markiewicz, Jeremy Mock, Tonee Smith, Natalia Toro, and many others.  

Contributions from Stanford and the University of Virginia are supported by the US Department of Energy under grants DE-SC0022083 and  DE-SC0007838, respectively. Elizabeth Berzin is supported by the National Science Foundation Graduate Research Fellowship under Grant No. DGE-2146755, and by the Stanford Graduate Fellowship. Jessica Pascadlo's work was supported by the U.S. Department of Energy, Office of Science, Office of Workforce Development for Teachers and Scientists, Office of Science Graduate Student Research (SCGSR) program. The SCGSR program is administered by the Oak Ridge Institute for Science and Education (ORISE) for the DOE. ORISE is managed by ORAU under contract number DE- SC0014664. SLAC author contributions are supported by Stanford University under Contract No. DE-AC02-76SF00515 with the U.S. Department of Energy, Office of Science, Office of High Energy Physics.  Andrew Whitbeck's and Joesph Kaminski's contributions are supported by DE-the Department of Energy under DE-FOA-0002112. All opinions expressed in this paper are the author’s and do not necessarily reflect the policies and views of the NSF, DOE, ORAU, or ORISE.

LKB acknowledges support from the Royal Swedish Academy of Sciences and the Knut and Alice Wallenberg Foundation. 

\bibliographystyle{IEEEtran}
\bibliography{ri.bib}

\end{document}

%% file: abstract.tex
The Light Dark Matter eXperiment (LDMX) is a proposed fixed-target missing momentum search for sub-GeV thermal relic dark matter. LDMX aims to probe thermal dark matter targets with $10^{16}$ electrons on target. Such an approach requires a high-repetition rate, low-current beam, with an average of one electron on target per event. These requirements are well-suited to the DArk Sector Experiments at LCLS-II (DASEL) facility, which will take advantage of the unused RF buckets between LCLS-II bunches to produce a well-defined low-current beam with a 26.9 ns bunch spacing. This document describes the results of a measurement of dark current in the Sector 30 transfer line (S30XL) of the LCLS-II beam, using a prototype of the LDMX trigger scintillator (TS) subsystem.

%% file: intro.tex
\section{Introduction}
\label{sec:intro}

\begin{figure}[h]
    \centering
    \includegraphics[width=0.6\linewidth]{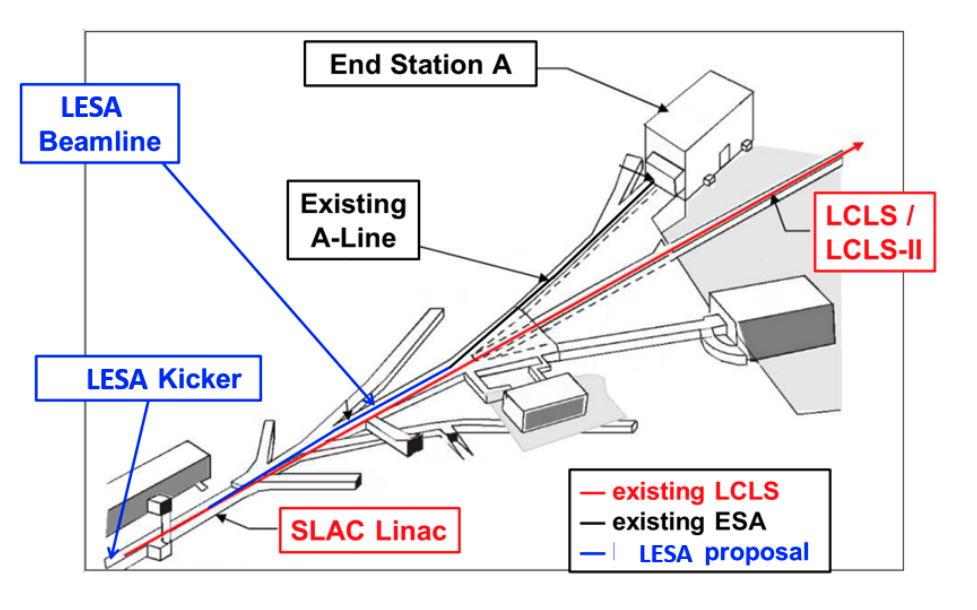}
    \captionsetup{width=0.89\linewidth}
    \caption{Layout of LESA and End Station A in relation to the SLAC Linac. Figure is from \cite{markiewicz2022slaclinacesalesa}.}
    \label{fig:LESA_layout}
\end{figure}

LDMX will operate at SLAC National Accelerator Laboratory (SLAC) in End Station A using the Linac to End Station A (LESA) beam facility. LESA will deliver electrons that are unused for generating x-ray pulses to End Station A, where LDMX will be installed. The layout of LESA is shown in Figure \ref{fig:LESA_layout}. The first stage of LESA construction and commissioning is the development of the Sector 30 Transfer Line (S30XL), a portion of the beamline that extracts electrons out of the LCLS-II beamline and into the A-Line. During the commissioning of S30XL, a prototype module of the LDMX trigger scintillator (TS) was installed in the beam line, with the goals of demonstrating successful parasitic running to LCLS-II, integrating the TS electronics with the LCLS-II timing system, and providing a time-resolved measurement of the LCLS-II dark current.

LCLS-II is fed by an RF gun operating at 186 MHz, the seventh subharmonic of the 1.3 GHz linac RF frequency. Free electron laser (FEL) pulses used for photon science occupy only a fraction of these available RF buckets, with a maximum bunch rate of 929 kHz \cite{markiewicz2022slaclinacesalesa}. This leaves 200 empty bunches in between FEL pulses, which are populated by dark current from the RF gun. The level of dark current was previously not known precisely, but was estimated to be less than 20 pA. 

\begin{figure}[h]
    \centering
    \includegraphics[width=0.7\linewidth]{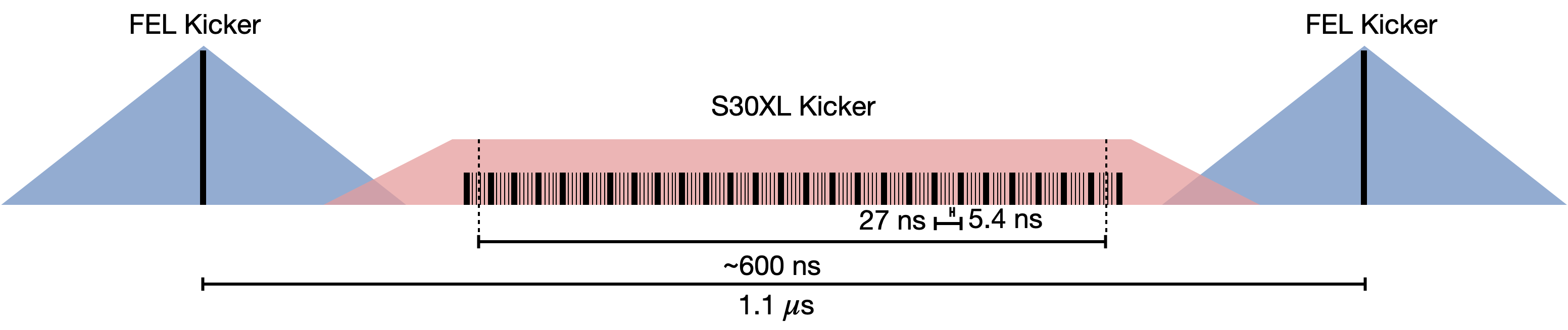}
    \captionsetup{width=0.89\linewidth}
    \caption{LCLS-II timing structure, showing primary FEL pulses, consisting of $10^8$ electrons, and bunches from the S30XL kicker at 37.14 MHz.}
    \label{fig:kicker_timing}
\end{figure}

The S30XL kicker is designed to divert these unfilled bunches down the S30XL beam line. Figure \ref{fig:kicker_timing} summarizes this timing structure, with the S30XL kicker flat-top spanning $\sim 600$ ns between FEL pulses arriving every 1.1 $\mu$s. For LDMX operations, a dedicated laser will be used to produce a well-defined, low-current beam at a bunch spacing of 37.14 MHz, with additional downstream spoilers and collimators used to further deplete out-of-time bunches relative to in-time bunches at the nominal beam rate. The thick vertical black lines in Figure \ref{fig:kicker_timing} denote these individual 37.14 MHz bunches, while thinner black lines show the 186 MHz RF frequency. A precise measurement of the LCLS-II dark current, in addition to being of general interest to the operation of LCLS-II, will therefore inform requirements and expectations for the LDMX laser and spoiler systems.

%% file: detector.tex
\section{Detector and experimental setup}
\label{sec:detector}

The detector components described in this section represent prototypes of systems that will be used in LDMX. 

\subsection{Trigger Scintillator}

\begin{figure}[h]
    \centering
    \includegraphics[width=0.5\linewidth]{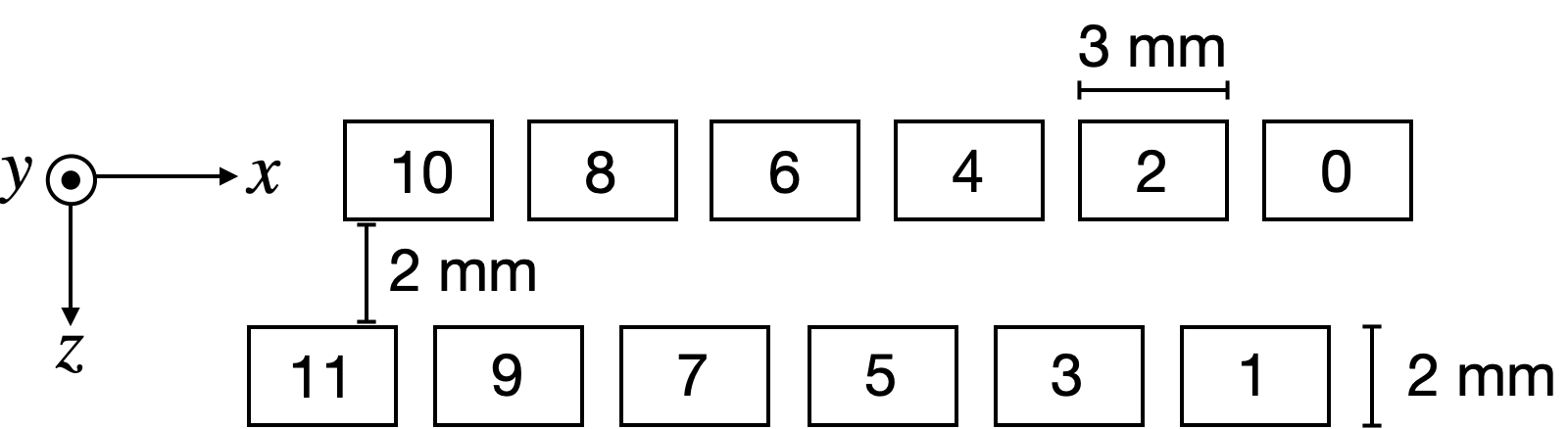}
    \captionsetup{width=0.89\linewidth}
    \caption{Configuration of scintillating bars in the trigger scintillator prototype installed in the S30XL beam line, with the beam direction along the $z$-axis.}
    \label{fig:bar_diagram}
\end{figure}

The TS module installed in the S30XL beam line consists of 2 $\times$ 3 $\times$ 30 mm EJ-200 polyvinyltoluene (PVT) bars, coupled to $2 \times 2$ mm$^2$ SiPMs. The module uses twelve bars arranged in two rows of six, with the long dimension along the $y$-direction, the 3 mm dimension along the $x$-direction, and the 2 mm dimension along the beam line ($z$-direction). The bar configuration and coordinate system is shown diagrammatically in Figure \ref{fig:bar_diagram}. The rows are staggered by half of a bar width, such that an electron passing through the gap of one layer will be contained within the other layer. 

The bars are arranged to have a $\sim0.3$ mm spacing in the $x$-direction, with the two rows separated by a $\sim2$ mm sheet of rubber. The bars in each row are separated by a sheet of enhanced specular
reflector film (ESR), and each row is sandwhiched by two long sheets of ESR. A light-tight housing holds the bars in this configuration, and interfaces with an array of 2 $\times\ 6$ SiPMs, arranged to map directly to the scintillators. The SiPMs are mounted on a single PCB, capable of supplying bias voltages and transmitting analog signals to the readout electronics.  

\subsection{SiPMs}

Hammamatsu S13360-2050VE SiPMs \cite{hamatsu:mppc}, with 1584 pixels in a $2\times2$ mm$^2$ cross-sectional area, were used for light detection. These SiPMs achieve peak sensitivity at 450 nm, and are therefore well matched to the scintillating bars, which have peak emission at a wavelength of 425 nm. Additional details about SiPM calibration and response are given in Section \ref{sec:SiPM_gains}.

\subsection{Readout Electronics}
The 12 SiPMs are read out using a CMS HCal Readout module, built on the QIE11 ASIC \cite{QIE}. Power and control signals are distributed to the readout module through a custom backplane. Each 37.14 MHz clock cycle, the  QIEs integrate the current signal from the SiPMs, outputting an 8-bit ADC and a 6-bit TDC. The QIE11 achieves deadtime-less digitization by running four operational phases in parallel, and has a dynamic range of $\sim400$ pC, with adjustable shunts at the input current that can further increase the range by a factor of 12. Given a typical SiPM gain of 200 fC/PE (photo-electron), readout is limited by saturation of the SiPMs, rather than by saturation of the QIE. 

The TDC provides the time within the sampling period at which the current pulse first exceeds a configurable current threshold. The 26.9 ns sampling period is sub-divided into 50 bins, providing 0.538 ns timing resolution. In sampling periods where the current never rises above threshold, the TDC is reported as 63; in periods where the current was above threshold at the start, the TDC is reported as 62.

\subsection{Timing and Control}
The TS readout electronics require a beam-synchronized clock (in order to align TDC and ADC integration boundaries with the beam), as well as other synchronization signals to align data streams on different fibers. This is achieved using a custom board based on a Zynq FPGA, referred to as a zynq Clock and Control Module (zCCM). The zCCM recovers the 186 MHz LCLS-II reference clock from a fast control signal described in the next section, and delivers a 37.14 MHz clock to the TS front-end (the fifth subharmonic of the LCLS reference clock). 

The zCCM also manages slow control signals, including enabling of power and I2C interfaces used for configuring the QIEs. A mezzanine board on the zCCM provides an interface to the TS front-end, designed to robustly deliver both fast and slow control signals over large distances. 

\subsection{Triggering and Data Acquisition}

An Advanced Processor (APx) board developed for the HL-LHC upgrade of the CMS experiment is used for triggering and data processing. The APx also houses a timing hub that decodes LCLS-II timing information and transmits timing messages to the zCCM. 

Data is transmitted from the readout module to the APx through an optical link. Following a positive trigger decision, the APx transmits a ReadOut Request (ROR) to the zCCM. For each ROR, a configurable number of 37.14 MHz samples both before and after the sample that initiated the trigger decision are recorded, for up to 128 total samples.

Three triggering approaches are implemented: a threshold trigger, a synthetic trigger, and a kicker trigger. The threshold trigger uses a simple High-Level Synthesis (HLS) algorithm to identify whether the charge measured in any TS bar is above a configurable PE threshold. The threshold trigger is used for characterization of backgrounds potentially uncorrelated to beam timing, as well as cosmic triggering. The synthetic trigger sends RORs at a configurable frequency in a configurable pattern, and is used primarily for SiPM characterization. The kicker trigger is synchronized to the 10 Hz signal in LCLS-II timing message, in time with the firing of the kicker.

Data received from the readout module following a ROR is then parsed and inserted into an Sqlite database. Each time sample that is read out includes the timestamp of the ROR, the ADC and TDC information from the QIEs for each channel, as well as data quality monitoring information from the QIEs, for identification of potential misalignment between channels or data transmissionfibers.

\subsection{Mechanics}

The TS module and supporting front-end electronics, including the QIE-based readout module and backplane, were installed in a region of the S30XL beam line immediately before the S30XL dump. 

\begin{figure}[h]
  \centering
  \captionsetup{width=0.89\linewidth}
  \begin{subfigure}[t]{0.56\textwidth}
    \includegraphics[width=\textwidth, height=5.5cm]{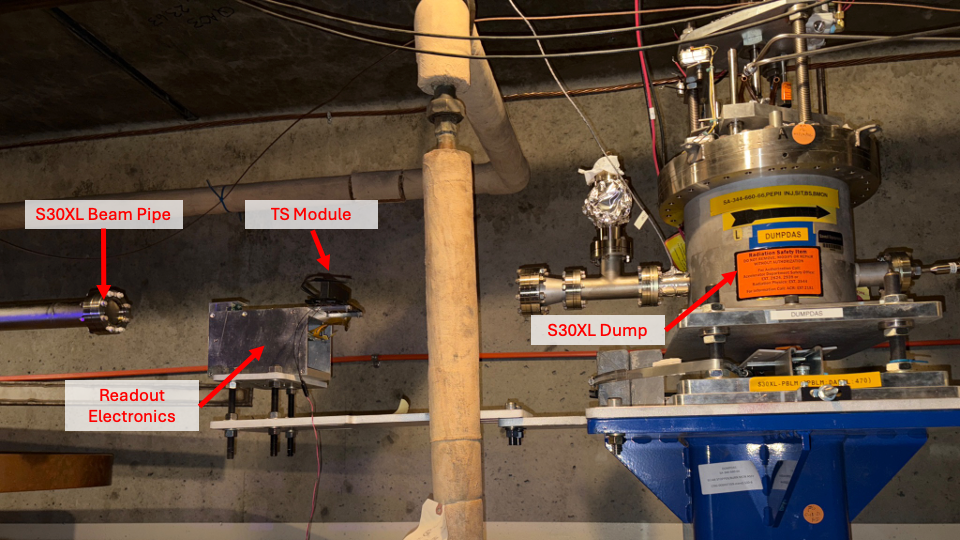}
    \caption{}
    \label{fig:main}
  \end{subfigure}%
  \hspace{0.6cm}
  \begin{subfigure}[t]{0.27\textwidth}
    \includegraphics[width=\textwidth,height=5.5cm]{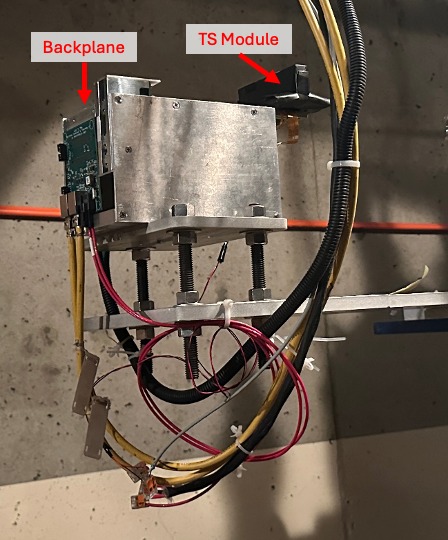}
    \caption{}
    \label{fig:small}
  \end{subfigure}
  \caption{TS module installed in the S30XL beam line, with S30XL beam pipe, readout electronics housing, and the S30XL dump indicated. A closer view of the TS and electronics housing is shown on the right, with the backplane labeled.}
  \label{fig:TS_photos}
\end{figure}

Figure \ref{fig:TS_photos} shows the positioning of the TS in the beam line. As indicated in Figure \ref{fig:TS_photos}, the front-end electronics, including the readout module and backplane, are housed in an aluminum support box below the beam line. Cables supplying low-voltage power to the readout electronics, SiPM bias voltages, slow and fast control signals from the zCCM, and data links from the readout module extend from the front-end housing to an electronics rack located above the beam tunnel. 

Before the run period, the center of the instrumented TS module was aligned with the center of the S30XL beam pipe. Along with the capability to additionally steer the beam spot by up to $\sim1$ cm in any direction, this ensured that the full beam spot would be contained by the scintillating bars.

%% file: results.tex
\section{Results}
\label{sec:results}


\subsection{SiPM Calibration} \label{sec:SiPM_gains}

\begin{figure}[h]
    \centering
    \captionsetup{width=0.89\linewidth}
\includegraphics[width=0.9\textwidth]{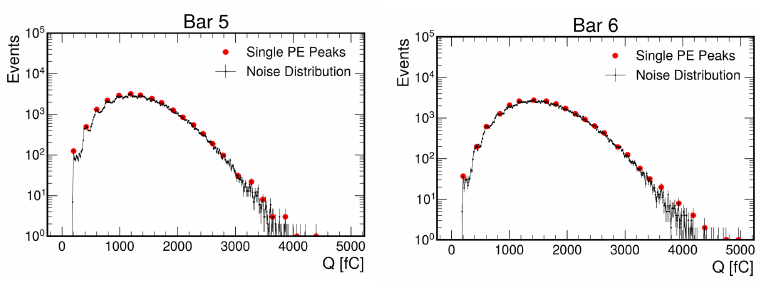}
    \caption{Distributions of charge integrated over 10 samples for two TS bars, during a period with no active dark current. Distinguishable peaks corresponding to individual photoelectrons are identified by red points.}
    \label{fig:pedestal_dists}
\end{figure}

SiPM gains and pedestals were measured and recorded for each data-taking period. Figure \ref{fig:pedestal_dists} shows the distribution of integrated charge over a ten sample window for each channel, at a bias voltage of 53 V (corresponding to a 3 V over-voltage). The red points denote peaks corresponding to individual numbers of photoelectrons. Peaks are found by selecting the largest sample (relative to neighboring samples) within a moving seven sample window. This window size was informed by expectations for single PE resolution. We note that the TS module is affected by a light leak, causing a broadening of the noise distribution. Despite this, the noise distribution remains well-separated from signal, expected to appear at $\sim$16000 fC.

\begin{figure}[h]
    \centering
    \captionsetup{width=0.89\linewidth}
\includegraphics[width=0.9\textwidth]{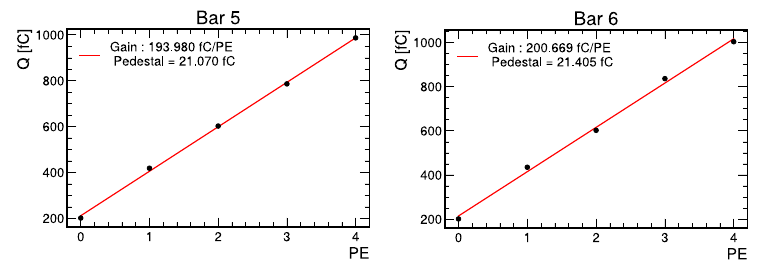}
    \caption{Charge of individual PE peaks (as determined by the first five points in Figure \ref{fig:pedestal_dists}) as a function of PE index, with linear fits for each channel. The slopes and $y$-intercepts of these fits give SiPM gains and pedestals, respectively.}
    \label{fig:gain_fits}
\end{figure}

Figure \ref{fig:gain_fits} shows linear fits to the positions of the first five peaks in Figure \ref{fig:pedestal_dists}, where the slope and $y$-intercept of the fit represent the gain and pedestal (over the number of integrated samples), respectively. The SiPMs are found to have gains of 180-200 fC/PE, or 1.2 $\times$ $10^6\ e^-$/PE , with pedestals of $\sim$20 fC. This measured gain is consistent with expectations for these SiPMs.

\subsection{Dark Current Measurement}
Dark current measurements were taken with the S30XL kickers configured to divert dark-current bunches down the S30XL line at a rate of 10 Hz. The LDMX TS module was triggered on the 10 Hz LCLS-II timing signal, and configured to read out 128 samples (corresponding to a window of 3.44 $\mu$s) per trigger. Additional runs were taken with the S30XL kickers disabled in order to characterize noise rates during the dark current window. This background rate was found to be negligible, with less than one MIP-like signal observed over the full data-acquisition window in a run length of one hour, suggesting a background current on the order of 20 aA. 

\begin{figure}[h]
    \centering
    \captionsetup{width=0.89\linewidth}
\includegraphics[width=0.9\textwidth]{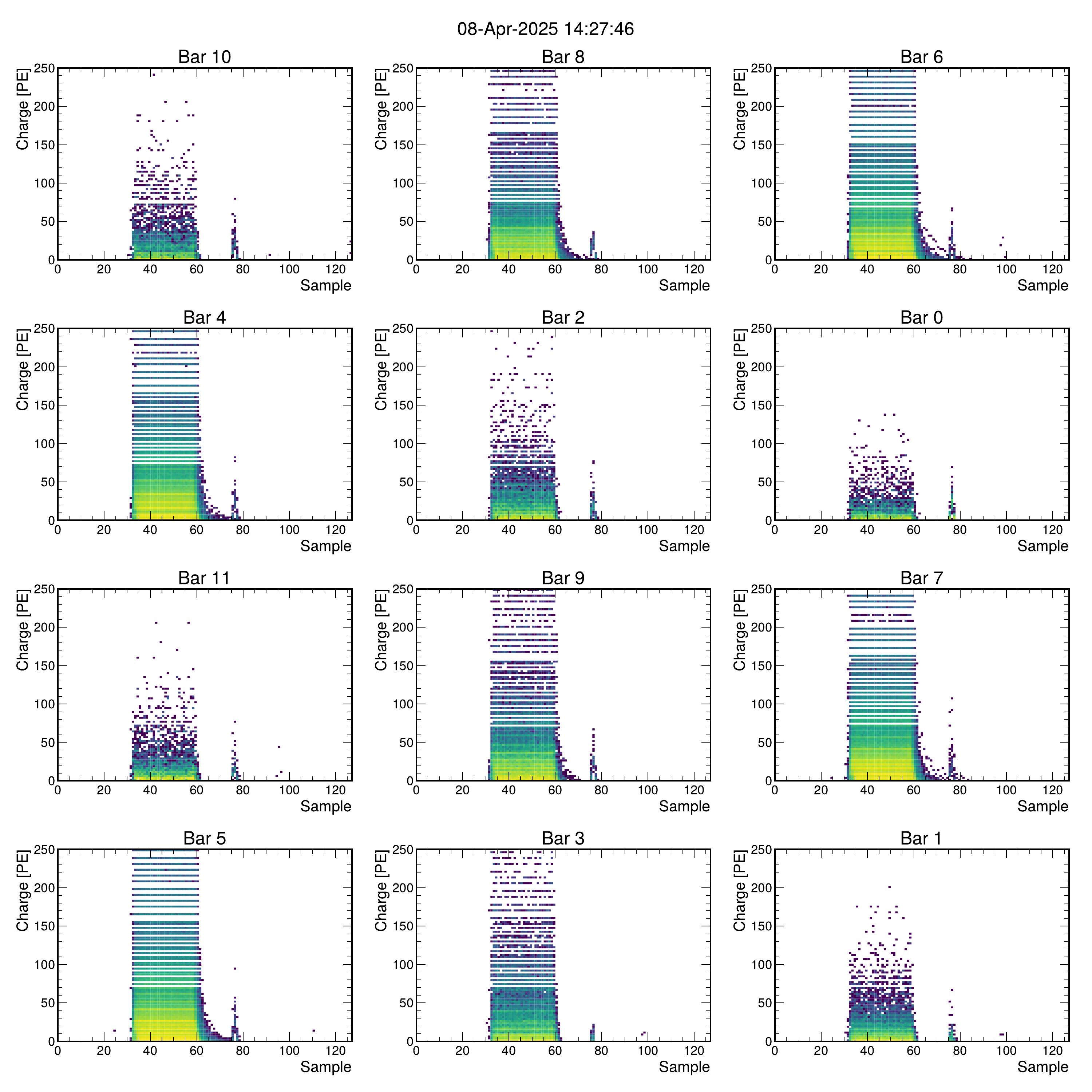}
    \caption{Accumulated waveforms over 5.5 hours of running for each TS channel, triggering on the LCLS-II timing signal with S30XL kickers enabled. The horizontal axis denotes time sample within the readout window, at a sampling frequency of 37.14 MHz.}
    \label{fig:charge_v_samples}
\end{figure}

Figure \ref{fig:charge_v_samples} shows accumulated charge waveforms in each TS bar for 5.5 hours of data acquisition. The period of near-continuous activity between time samples 30 and 60 is characteristic of a dark current signal, in agreement with the expected duration of the S30XL kicker flat-top. An additional pulse can be seen at sample 75 in Figure \ref{fig:charge_v_samples}. This pulse was intermittently observed in waveforms with and without the S30XL kickers enabled, and is likely an electron from upstream activity in the LCLS-II line.

\begin{figure}
    \centering
    \captionsetup{width=0.89\linewidth}
    \includegraphics[width=1\linewidth]{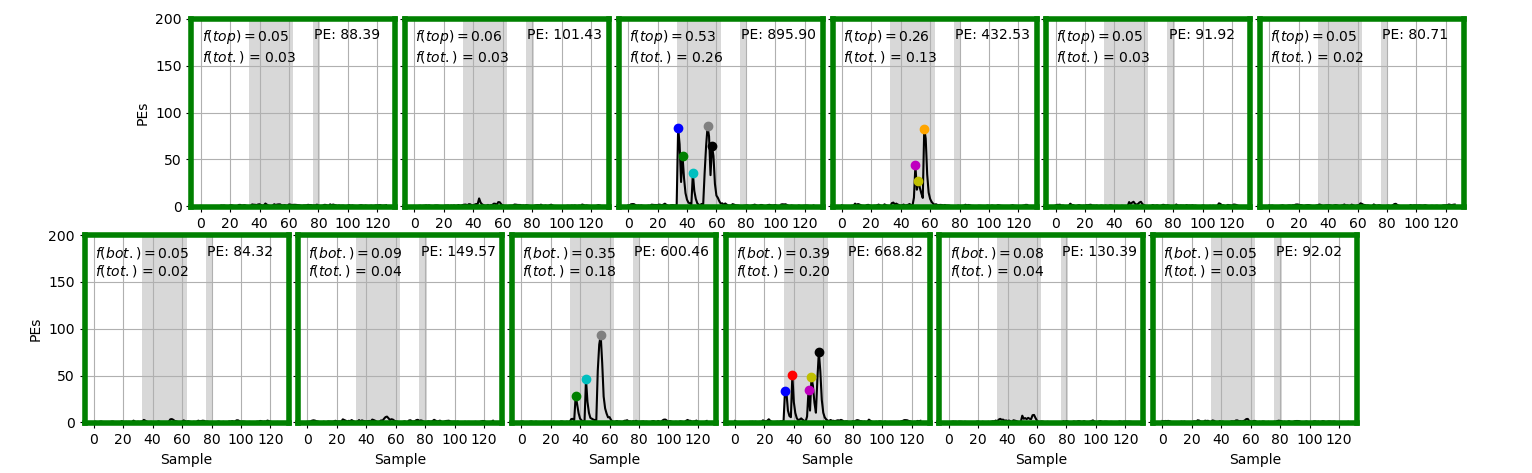}
    \caption{Display of measured waveforms from a single representative event when the beamline was transporting dark current. The colored points indicate local maxima in the PE distribution.  Those points with the same color arrive in the same time sample.  The large gray shaded region indicates the kicker flat top.  The small gray region indicated a period of high background. The plots are arranged in the same physical orientation as the scintillator bars.}
    \label{fig:event_example}
\end{figure}

An example of a typical waveform for an individual event is shown in Figure \ref{fig:event_example}, with grey bands denoting the kicker window (corresponding to the observed periods of dark current activity in Figure \ref{fig:charge_v_samples}) and the background pulse at sample 75. Individual electron peaks are marked by colored points, where points with the same color indicate electrons that arrived within the same 27.6 ns bunch. 

\begin{figure}[h]
    \centering
    \captionsetup{width=0.89\linewidth}
\includegraphics[width=0.5\textwidth]{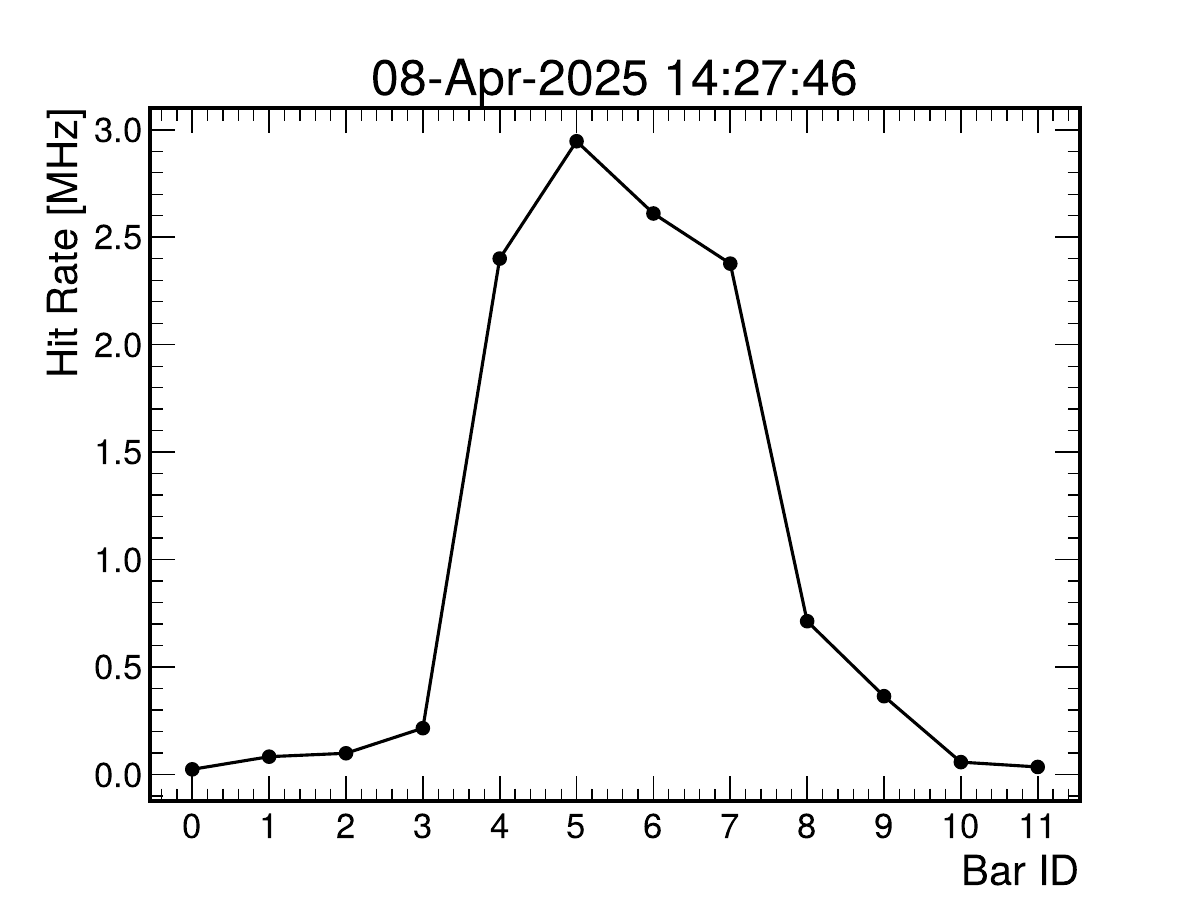}
    \caption{Rate of hits exceeding a 20 PE threshold (within the kicker flat-top window) as a function of bar number.}
    \label{fig:beam_profile}
\end{figure}

Figure \ref{fig:beam_profile} shows the rate of hits within the kicker window above a 20 PE threshold as a function of bar number. This threshold is well-separated from the noise distribution plotted in Figure \ref{fig:pedestal_dists}, which suggests $< 1$ PE from SiPM noise per sample, and is well below the value of a single MIP.

We see that this rate approaches zero near the edges of the TS module, suggesting that the beam spot is fully contained within the area spanned by the scintillator.  This was further confirmed through minor adjustments to the position of the beam spot before the start of the run period. Steering of the beam in the horizontal direction (in the direction of increasing/decreasing bar number) suggested that Figure \ref{fig:beam_profile} reflects the full beam profile. No change in hit rate was observed when steering the beam by $\sim \pm 1$ cm in the vertical direction (along the 30 mm length of the TS bars). Given that the 30 mm length of the TS spans the majority of the beam pipe, and the expected diameter of the beam spot is on the order of millimeters, this suggests that the TS fully contains the beam in the vertical direction as well as the horizontal direction. 

\begin{figure}[h]
    \centering
    \captionsetup{width=0.89\linewidth}

\includegraphics[width=0.9\textwidth]{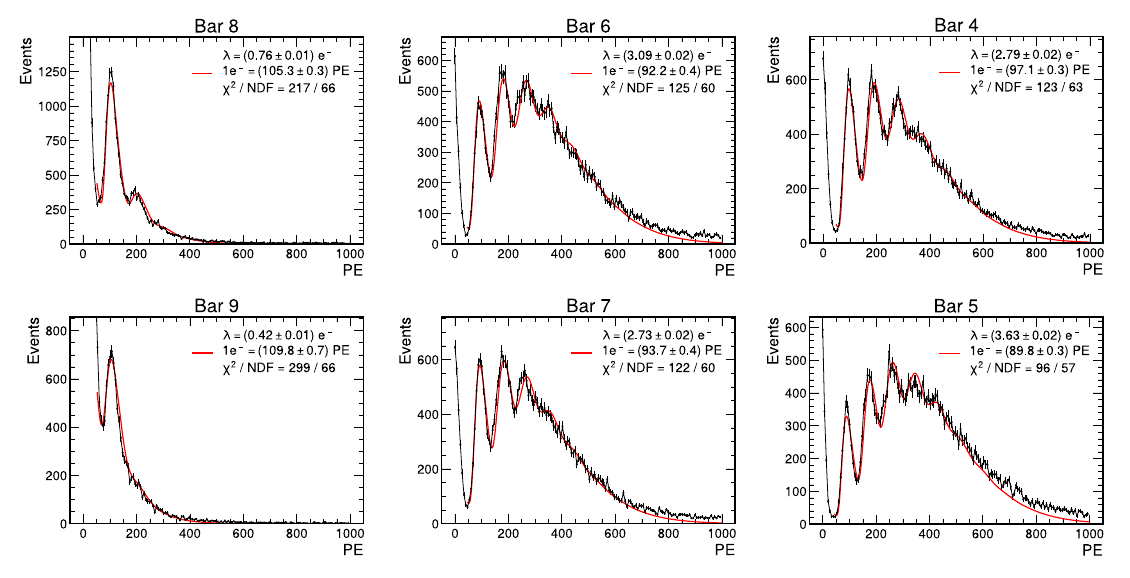}
    \caption{Total charge in PEs integrated over the kicker flat-top window when the beamline was transporting dark current. PE distributions in bars within the beam spot are fit to a Poisson-distributed sum of Landau peaks (red line). $\lambda$ indicates the Poisson average number of electrons for each bar. The single-electron light yield for each bar within the beam spot is also indicated.}
    \label{fig:charge_hists}
\end{figure}

The level of dark current is estimated using two complementary approaches. The first approach uses the total charge measured within each kicker window, found by integrating each waveform between time samples 30 and 60. These charge distributions are shown for each channel in Figure \ref{fig:charge_hists}. The red lines show fits to a Poisson-distributed sum of Landau peaks:
\begin{equation}
    f_Q(n_{\text{PE}}) = A \sum_{n=1}^{5} \text{Poisson}(n;\lambda) \times \text{Landau}(n_{\text{PE}};nQ_{\text{MIP}},\sigma\sqrt{n}),
\end{equation}
where $A$ is a scaling factor, $\lambda$ is the average number of electrons in the window, and $Q_{\text{MIP}}$ is the charge corresponding to a single electron, in PEs, and the width of the Landau peak is scaled according to the number of electrons $n$. We find that this model tends to underestimate the high-charge tails of these distributions, indicating non-Poissonian behavior in the dark current. Summing total charge in each layer of the TS, we find $\lambda_{\text{tot}} \approx 6.6$-$6.7\ e^-$ per kicker flat top window, or about 0.05 (0.25) $e^-$ per 5.4 (26.7) ns bunch . Considering the $\sim$700 ns duration of the kicker flat-top, this corresponds to a dark current of $\sim1.5$ pA.

\subsection{Electron Counting}

Alternatively, as can be seen in Figure \ref{fig:event_example}, the relatively low level of dark current allows for the identification of individual electrons and their arrival times. Thus, an electron counting approach, similar to the strategy that will be used to count electrons using the full LDMX TS, can be used to estimate dark current.

Peaks were tagged with a threshold of 20 PE, and grouped into clusters depending on their arrival time and physical position. A ``cluster" was defined as a group of peaks in adjacent bars (according to the numbering scheme in Figure \ref{fig:bar_diagram}) with identified peak times in the same 26.9 ns sample. Cluster sizes were limited at two bars, given the geometric infeasibility of seeing legitimate three-bar clusters from electrons at normal incidence. We note that in rare cases, this approach may under-count electrons in the event that more than one distinct electron arrives within the same 37.14 MHz clock cycle. Given a dark current rate of $\sim0.01\ e^-$/ns, as taken from Figure \ref{fig:charge_hists}, the probability of seeing more than one electron in the same 26.9 ns window is $\sim 2.7\%$, suggesting that the impact of such effects should be minimal.

\begin{figure}[h]
    \centering
    \captionsetup{width=0.89\linewidth}
\includegraphics[width=0.5\textwidth]{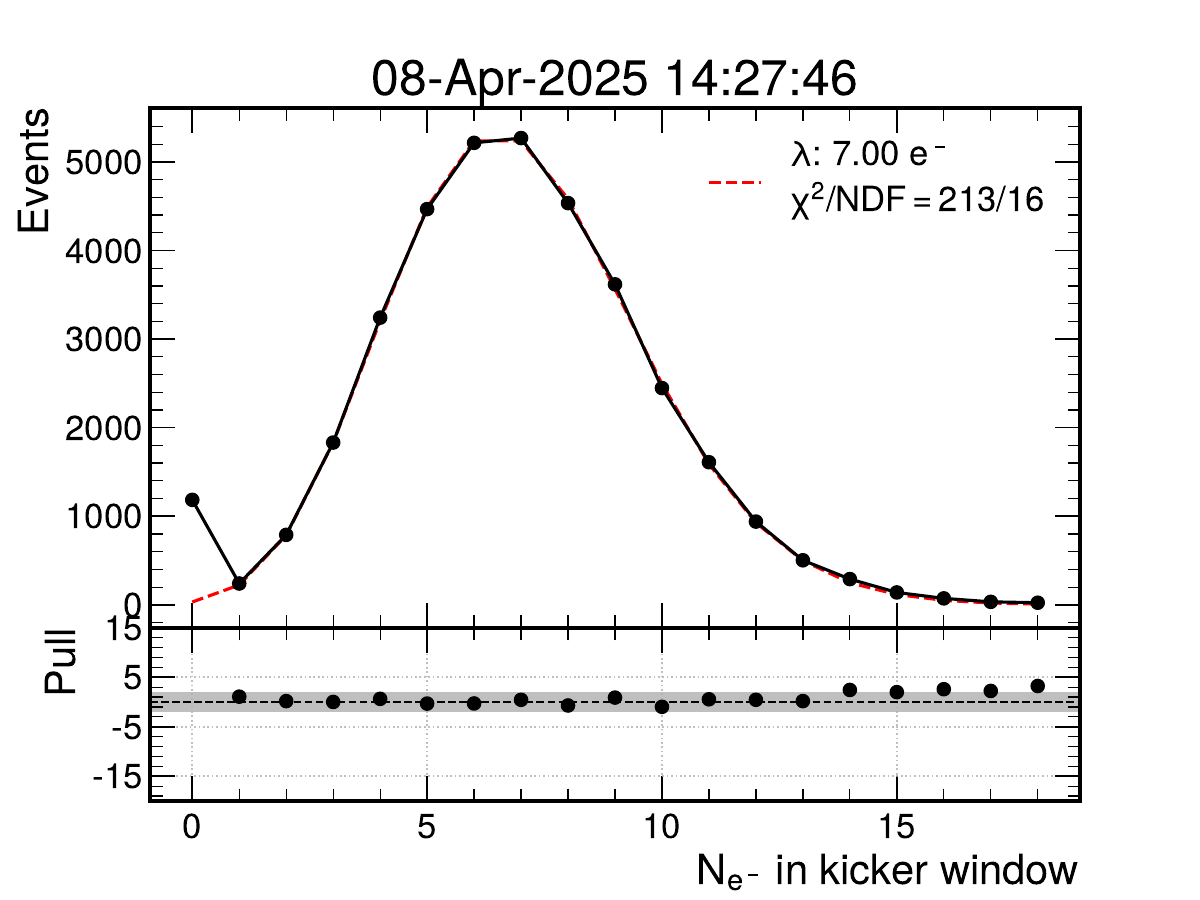}
    \caption{Distribution of the total number of electrons in the kicker window when the beamline was transporting dark current. The distribution is fit with a Poisson distribution with $\lambda = 6.98$.  The bottom panel shows the pulls (data - fit)/(error) from the fit with a $2\sigma$ band in gray .}
    \label{fig:electron_count}
\end{figure}

The distribution of the number of counted electrons per kicker flat-top window is shown in the top panel of Figure \ref{fig:electron_count}, with scaled residuals from a fit to a Poisson distribution in the bottom panel. We find that this electron-counting approach gives $\lambda_{\text{tot}} \approx 7.00\ e^-$, in agreement with the estimate directly from the charge distributions in Figure \ref{fig:charge_hists}. The increase in events with $N_{e^-} = 0$ is a result of intermittent interruptions to the S30XL kicker caused by other LCLS-II activity, during which no dark current was delivered to the prototype TS module. We also see a slight excess in the tail of the distribution at high $N_{e^-}$, consistent with the disagreement in the tails observed in Figure \ref{fig:charge_hists}.

\subsection{Timing Structure}

\begin{figure}[h]
  \centering
  \begin{subfigure}[t]{0.5\textwidth}
    \includegraphics[width=\textwidth, height=6cm]{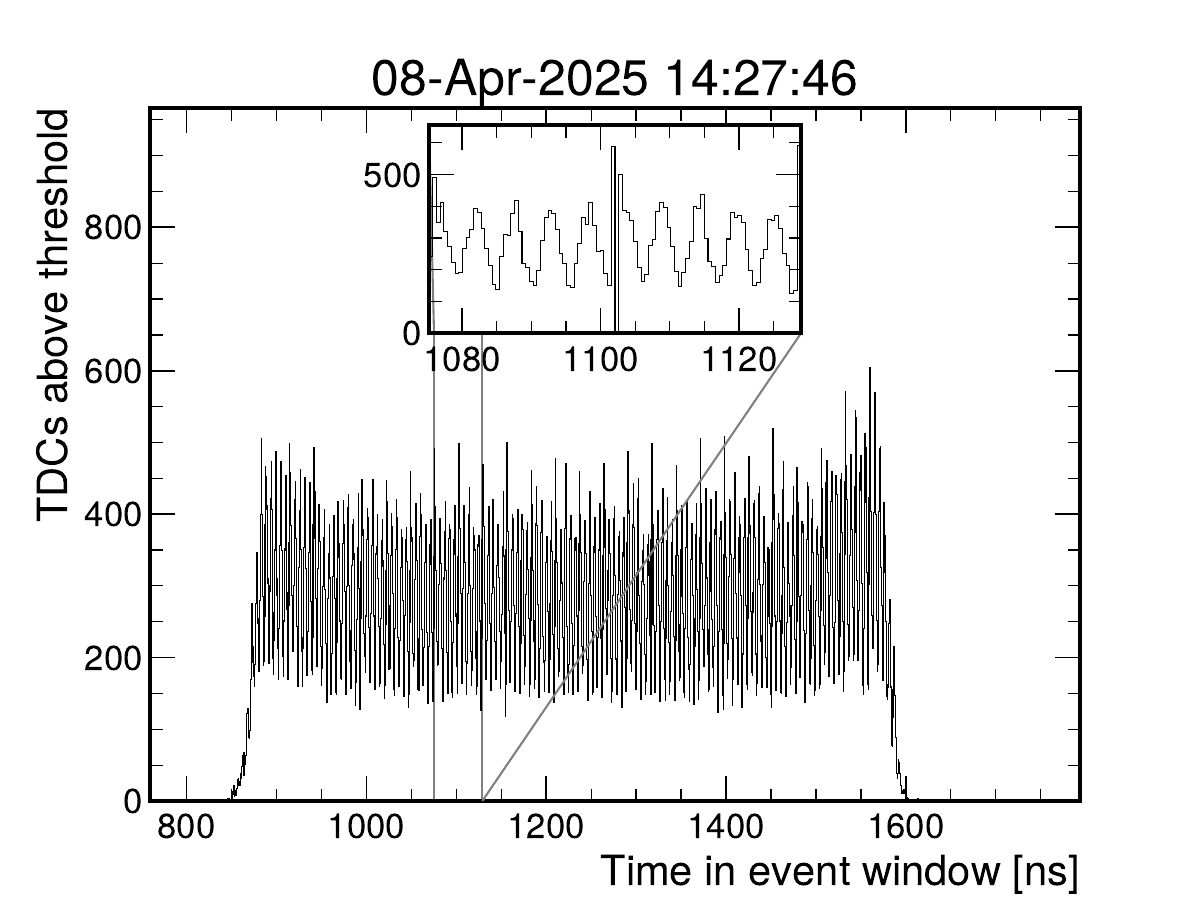}
    \caption{}
    \label{fig:all_tdcs}
  \end{subfigure}%
  \begin{subfigure}[t]{0.5\textwidth}
    \includegraphics[width=\textwidth,height=6cm]{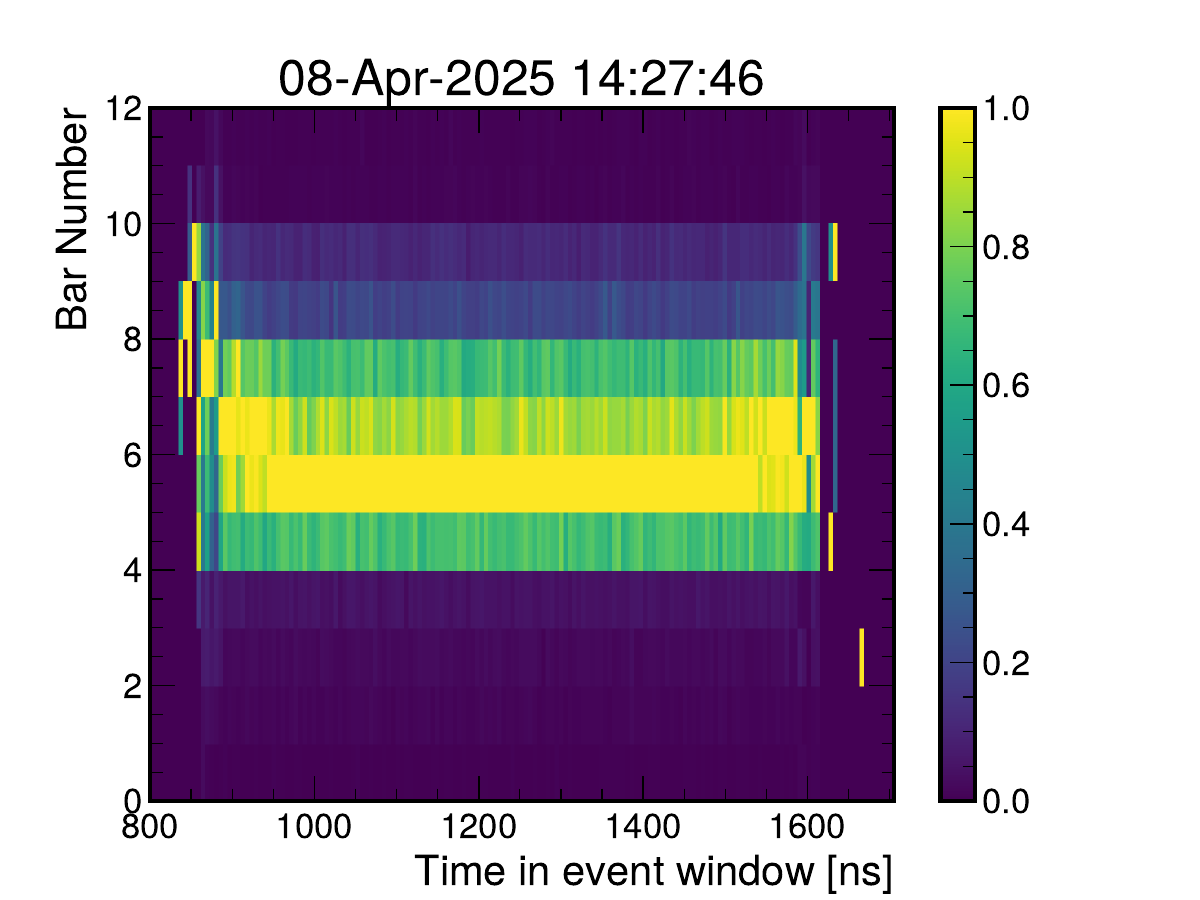}
    \caption{}
    \label{fig:time_v_bar}
  \end{subfigure}
  \caption{\textbf{Left:} Arrival times, in ns, of electron peaks in the kicker pulse window for the TS prototype installed in the S30XL beamline when the beamline was transporting dark current. Individual 5ns pulses can seen, corresponding to the 186MHz LCLS-II RF structure. \textbf{Right: } Change in beam position (in units of bar number) over time, showing the changing acceptance of the kicker magnet.
}
  \label{fig:TDC_structure}
\end{figure}

TDC information was used to measure precise electron arrival times. Figure \ref{fig:all_tdcs} shows the arrival times of the electrons with respect to the start of the readout window. We see clear peaks every $\sim5.38$ ns, corresponding to the 186 MHz RF gun frequency. The width of these peaks is dominated by the TDC timing resolution. In particular, this indicates that any potential smearing from the RF gun is subdominant. 

Figure \ref{fig:time_v_bar} shows the change in beam position (in units of bar number) over time. The changing acceptance of the kicker magnets over their $\sim150$ ns ramp can be seen in the change of the beam position at the beginning and end of the  window. The beam spot stabilizes for $\sim570$ ns, in agreement with the expected timing of the S30XL kicker, as depicted in Figure \ref{fig:kicker_timing}.

\subsection{Dark Current Stability}

\begin{figure}[h]
  \centering
  \begin{subfigure}[t]{\textwidth}
    \includegraphics[width=\textwidth]{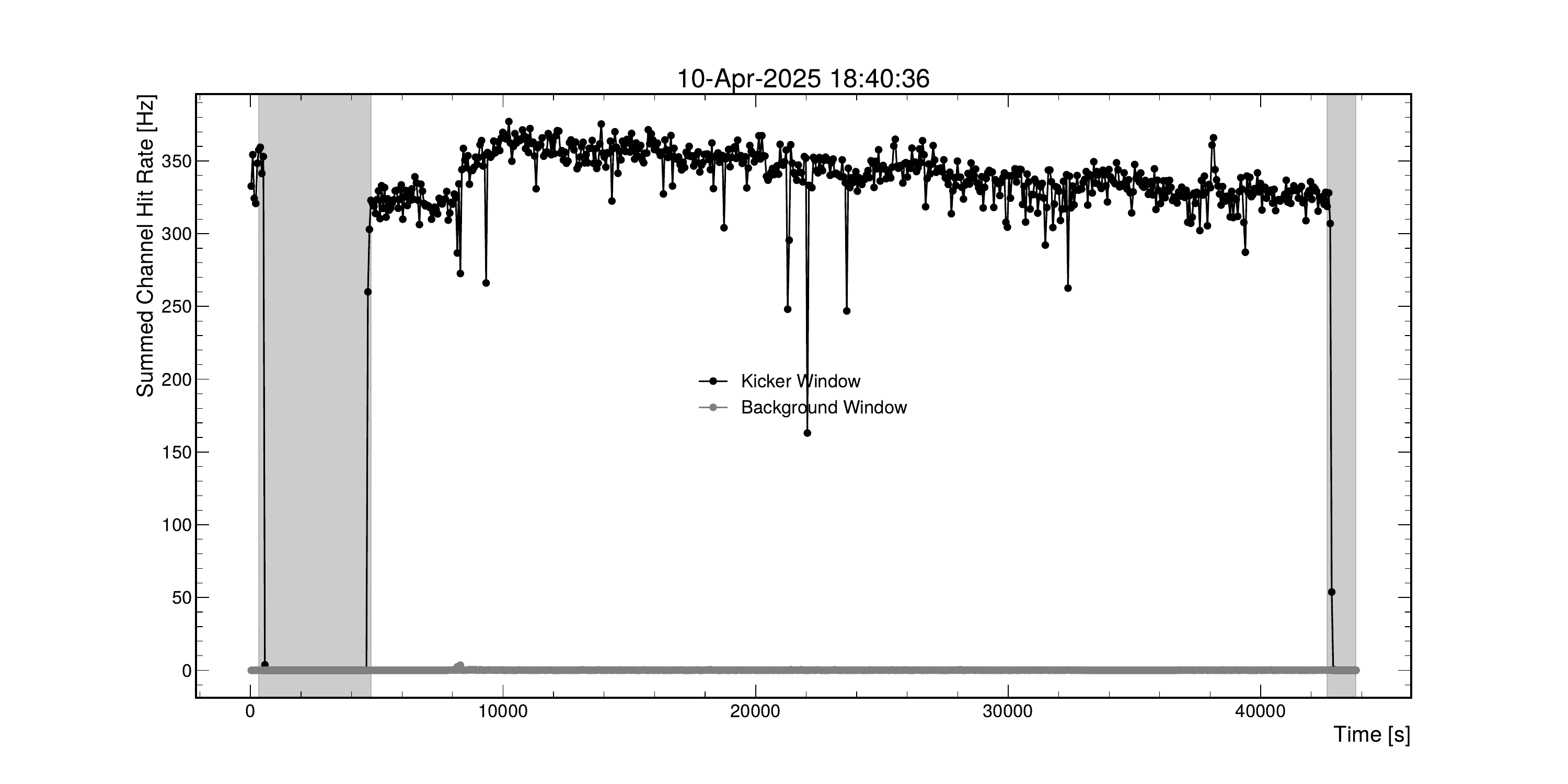}
    \caption{}
    \label{fig:summed_rate_0410}
  \end{subfigure}%
  \hfill
  \begin{subfigure}[t]{\textwidth}
    \includegraphics[width=\textwidth]{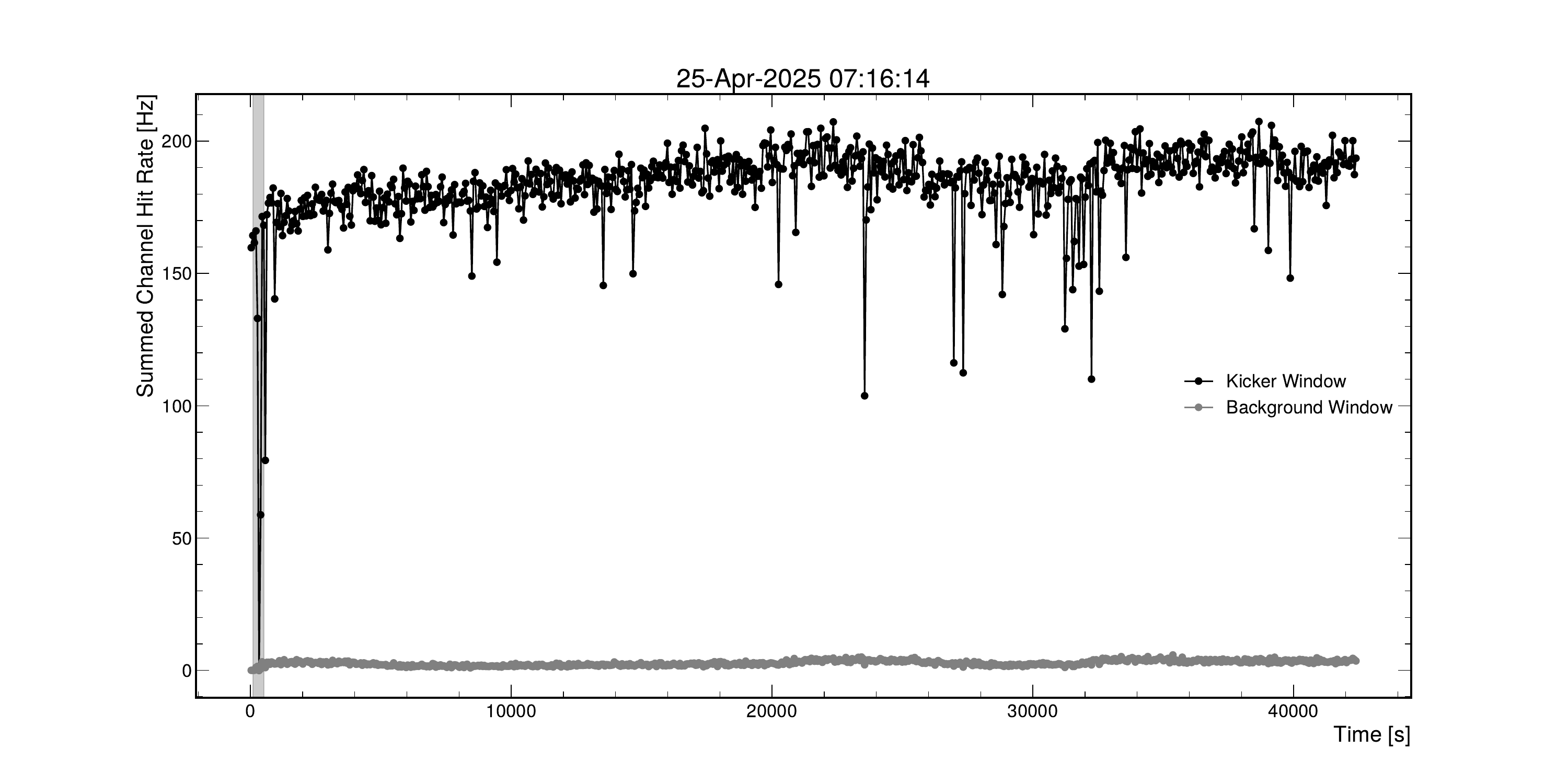}
    \caption{}
    \label{fig:summed_rate_0425}
  \end{subfigure}
  \caption{Rate of hits exceeding a 20 PE threshold (in black points), summed across all TS bars, for two different runs. Shaded gray regions indicated extended periods when dark current delivery to the S30XL dump was interrupted.}
  \label{fig:summed_rates}
\end{figure}

Dark current in the S30XL beam line was measured intermittently in 6-8 hour runs spanning a month-long period. Figure \ref{fig:summed_rates} shows the rate of samples exceeding a 20 PE threshold, summed across all channels, for two of these runs. This rate is a measure of dark current activity, and is consistent with the dark current determined through electron-counting--each electron is measured in multiple samples, and will tend to be seen in two channels.  We see that apart from occasional periods where the S30XL kickers were temporarily disabled (indicated in the shaded gray regions), the dark current rate within each run is fairly stable. 

\begin{figure}[h]
  \centering
  \begin{subfigure}[t]{0.32\textwidth}
    \includegraphics[width=\textwidth]{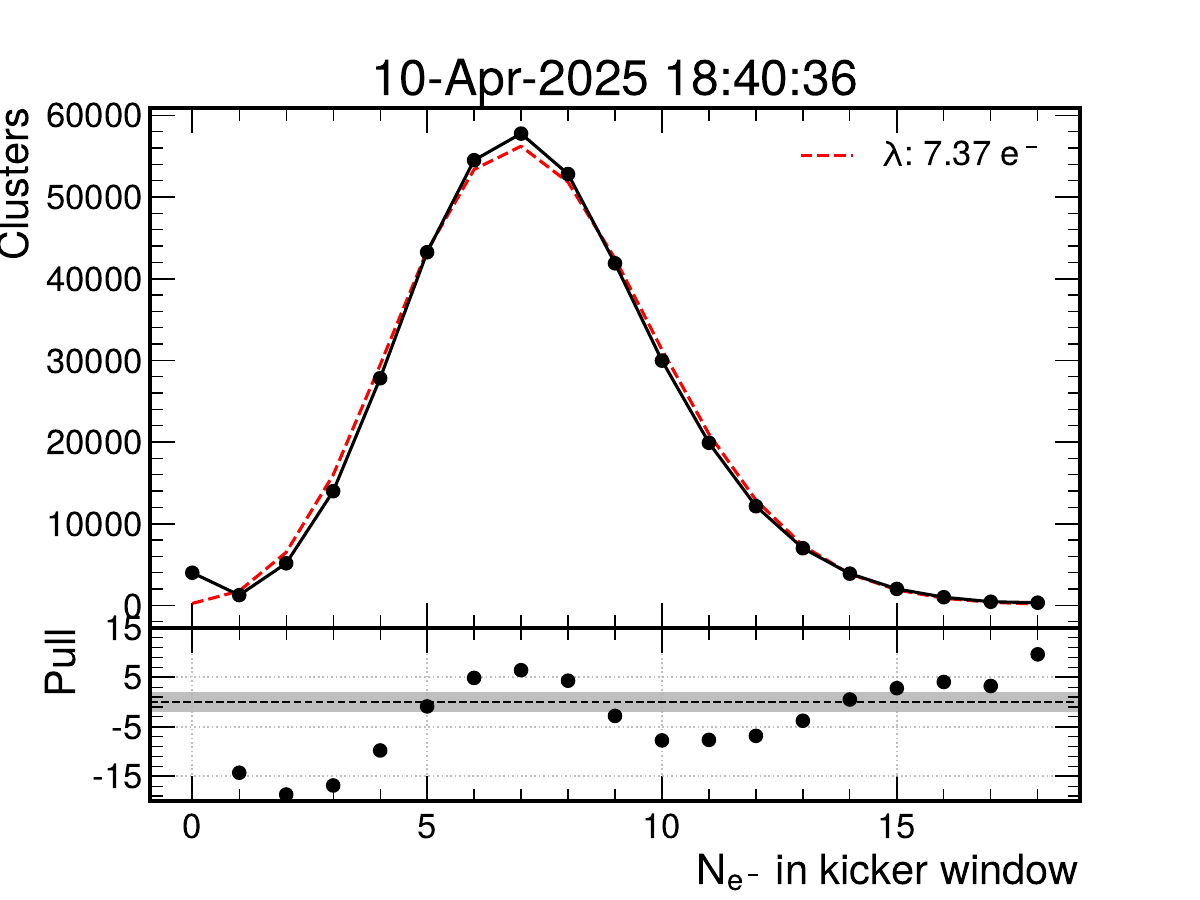}
    \caption{}
    \label{fig:ele_count_0410}
  \end{subfigure}%
  \hspace{0.001cm}
  \begin{subfigure}[t]{0.32\textwidth}
    \includegraphics[width=\textwidth]{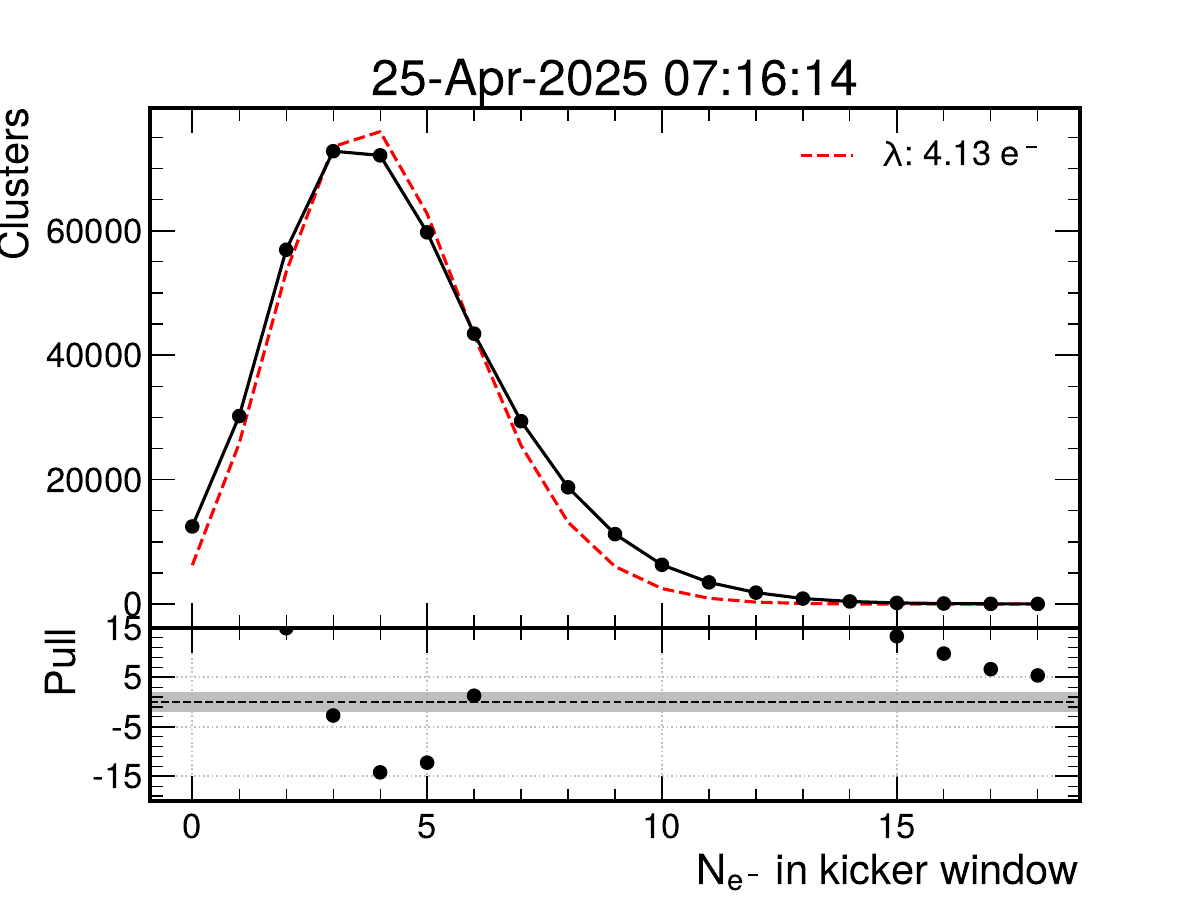}
    \caption{}
    \label{fig:ele_count_0425}
  \end{subfigure}
  \hspace{0.001cm}
  \begin{subfigure}[t]{0.32\textwidth}
    \includegraphics[width=\textwidth]{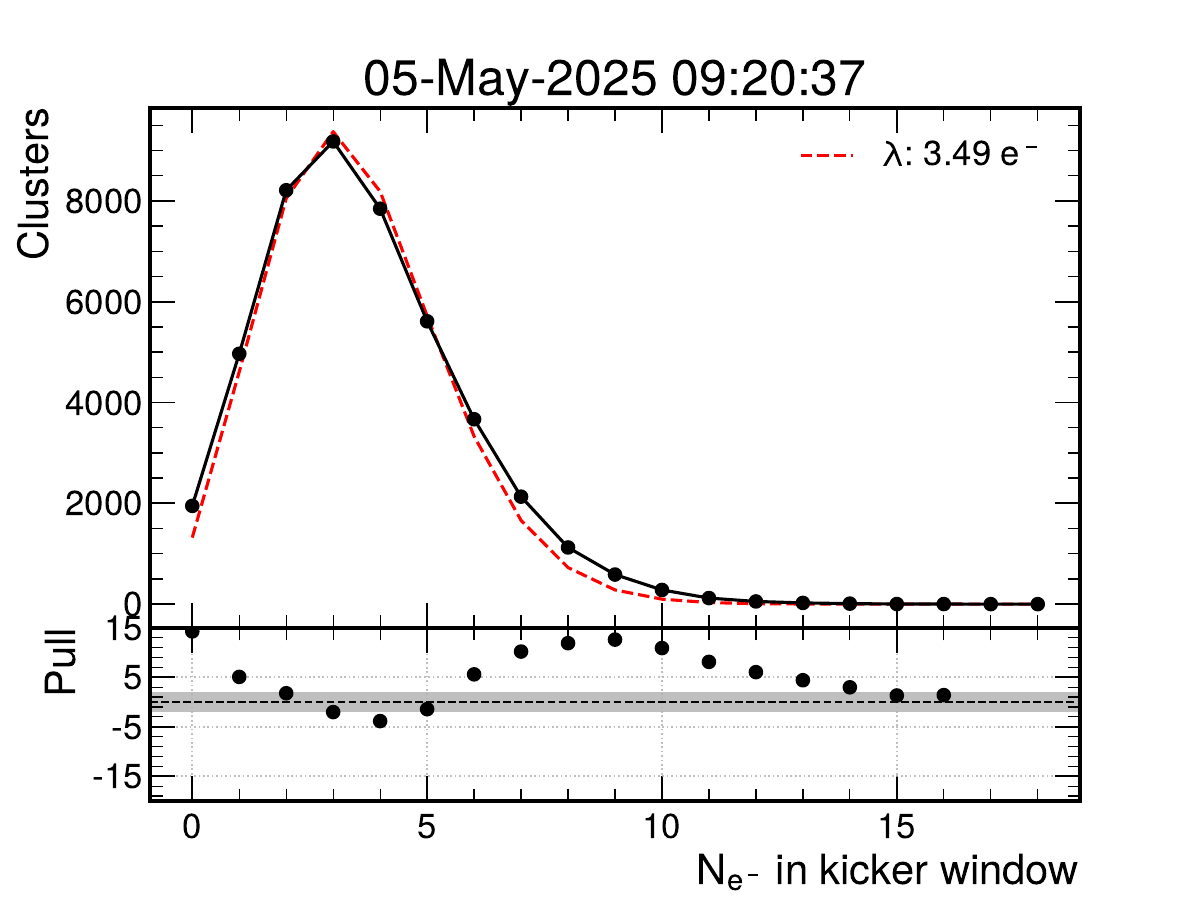}
    \caption{}
    \label{fig:ele_count_0505}
  \end{subfigure}
  \caption{Distribution of the total number of electrons in the kicker window when the beamline was transporting dark current, for three different runs.}
  \label{fig:electron_count_runs}
\end{figure}

Comparisons between runs show occasional discrete changes to the level of dark current. Figure \ref{fig:electron_count_runs} shows the number of electrons within the kicker flat-top window for three different runs. We see that the dark current ranges from $3.49$ to $7.37\ e^-$ within the flat-top, corresponding to currents between 0.8 and 1.7 pA. Some of these distributions show systematic disagreement from the Poisson fit, due to dark current rates that evolve over time.

%% file: conclusion.tex
\section{Conclusions}
\label{sec:conclusions}
This work has presented the measurement of the dark current of the LCLS-II beam utilizing a prototype LDMX trigger scintillator module. We demonstrated parasitic operation and integration with the LCLS-II timing system. The prototype provided a time-resolved measurement of the LCLS-II dark current, and characterized the dark current utilizing two complementary methods: charge integration and direct electron counting. These analyses consistently found the dark current to range between 0.8 and 1.7 pA at the sector 30 transfer line.
